\newif\ifsc
\newif\iftreport
\sctrue 
\treporttrue 
\ifsc
\documentclass[12pt, draftclsnofoot, onecolumn]{IEEEtran}
\else
\documentclass[conference,10pt]{IEEEtran}
\IEEEoverridecommandlockouts
\fi

\usepackage{cite}

\usepackage[cmex10]{amsmath}
\usepackage{amssymb,latexsym}
\usepackage{epsfig}
\usepackage{color}
\usepackage{graphicx}
\usepackage{url}
\usepackage{bbm}
\usepackage{epstopdf}

\usepackage{pst-plot}
\usepackage{pstricks-add}
\usepackage{breqn}
\usepackage{algpseudocode}
\usepackage{algorithm}
\usepackage{psfrag}
\usepackage{mathtools} 

\usepackage{enumitem}

\ifsc
\else
  \def\BibTeX{{\rm B\kern-.05em{\sc i\kern-.025em b}\kern-.08em
    T\kern-.1667em\lower.7ex\hbox{E}\kern-.125emX}}
\fi

 \allowdisplaybreaks

\newcommand{\newoperator}[3]{\newcommand*{#1}{\mathop{#2}#3}}

\newcommand{\tr}{\operatorname{tr}}
\newcommand{\E}{\mathbb E}

\newoperator{\diag}{\mathrm{diag}}{\nolimits}
\newoperator{\rank}{\mathrm{rank}}{\nolimits}
\newoperator{\myperm}{\mathrm{perm}}{\nolimits}
\newoperator{\var}{\mathrm{var}}{\nolimits}

\mathchardef\myhyphen="2D

\newcommand{\xbf}{\boldsymbol x}

\newcommand{\qbf}{\boldsymbol q}
\newcommand{\pbf}{\boldsymbol p}

\newcommand{\Pbf}{\boldsymbol P}

\newcommand{\herm}{\mathrm{\dagger}}

\newoperator{\myvec}{\mathrm{vec}}{\nolimits}

\definecolor{myblue}{rgb}{0.3328,    0.3539,    0.7758}
\definecolor{myblue2}{rgb}{0.0328,    0.0539,    0.4758}
\definecolor{mygreen2}{rgb}{ 0.0328 0.4758    0.0539} 
\definecolor{mygreen3}{rgb}{ 0.0328 0.1758    0.0539} 
 \definecolor{myred}{rgb}{0.4758, 0.0328,    0.0539}

\algnewcommand{\Initialize}[1]{%
  \State \textbf{Initialize:}
  \Statex \hspace*{\algorithmicindent}\parbox[t]{.8\linewidth}{\raggedright #1}
}

\newcommand{\err}{\ensuremath{\varepsilon}} 

\newcommand{\ns}{\ensuremath{n_s}}

\newtheorem{prop}{\bf{Proposition}}[section]

\begin{document}

\title{Optimization vs. Reinforcement Learning for Wirelessly Powered Sensor Networks
\thanks{A.~\"Oz\c celikkale acknowledges the support from  Swedish Research Council under grant 2015-04011. M.~Koseoglu acknowledges the support from  Fulbright Program with grant number FY-2017-TR-PD-02.}
}

\author{\IEEEauthorblockN{Ay\c ca \"Oz\c celikkale\IEEEauthorrefmark{1},
Mehmet Koseoglu\IEEEauthorrefmark{2} and
Mani Srivastava\IEEEauthorrefmark{2} 
}\\
\IEEEauthorblockA{\IEEEauthorrefmark{1}Signals and Systems, Uppsala University, Uppsala, Sweden}\\
\IEEEauthorblockA{\IEEEauthorrefmark{2}Dept. of Electrical and Computer Engineering, University of California, Los Angeles, USA}
}

\date{}
\maketitle

 
 \begin{abstract}
We consider a sensing application where the sensor nodes are wirelessly powered by an energy beacon. We focus on the problem of jointly optimizing the energy allocation of the energy beacon to different sensors and the data transmission powers of the sensors in order to minimize the field reconstruction error at the sink. In contrast to the standard ideal linear energy harvesting (EH) model, we consider practical non-linear EH models. We investigate this problem under two different frameworks:  i) an optimization approach where the energy beacon knows the utility function of the nodes, channel state information and the energy harvesting characteristics of the devices; hence optimal power allocation strategies can be designed using an optimization problem and ii) a learning approach where the energy beacon decides on its strategies adaptively with battery level information and feedback on the utility function. Our results illustrate that deep reinforcement learning approach can obtain the same error levels with the optimization approach and provides a promising alternative to the optimization framework.
 \end{abstract}

\section{Introduction}
\iftreport
Wireless power transfer (WPT) is a promising technology for enabling energy-autonomous future networked systems \cite{Krikidis_2014,SchoberPoor_2015}. 
\fi
%
At the moment, a significant part of the literature on  WPT systems focus on linear energy harvesting (EH) models where the average power that can be harvested at the EH device is modeled as a linear function of the average power input to the device.
On the other hand, practical EH hardware circuitry design is limited by the non-linear characteristics of circuit components, which yields to energy harvesting efficiencies that highly depend on the input power levels and input wave-forms. 

Investigation of these issues have only recently started to appear in the communications community: 
\iftreport
Refs.~\cite{clerckxBayguzinaYatesMitcheson_2015,ClerckxBayguzina_2016,huangClerckx_2018, Clerckx_2018}  show the superior performance of multi-sine waveforms  for power transfer compared to the traditional communication waveforms.
\fi
Non-linear models for power conversion efficiency in EH circuitry are investigated and performance improvements due to usage of practical models in communication system design are illustrated \cite{BoshkovskaNgZlatanovSchober_2015,boshkovska_robust_2017,XuOzcelikkaleMcKelveyViberg_2017}. 
In this article, we contribute to this line of work by investigating the effect of non-linear power conversion on the performance of a  remote sensing system powered with WPT. 

\begin{figure}
\begin{center}
\includegraphics[width=0.80 \linewidth]{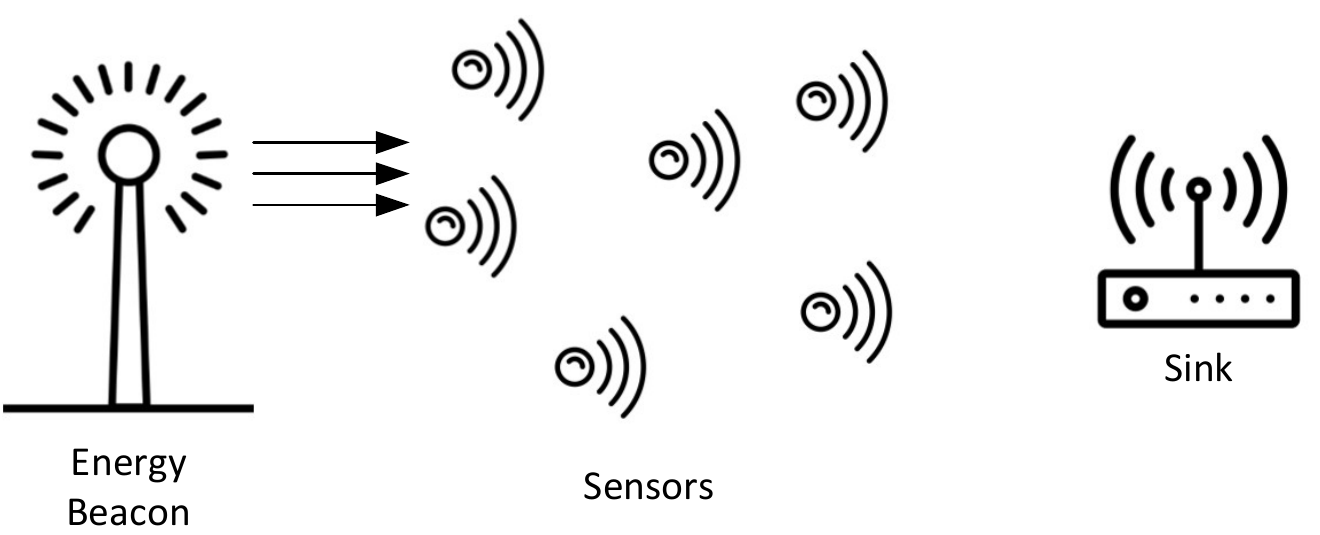}
\end{center}
\caption{Sensors powered by an energy beacon transmitting to a sink. }
\label{fig:scenario}
\kern-1em
\end{figure}

We consider the setting in Fig. \ref{fig:scenario} where the sensor nodes are wirelessly powered by an energy beacon. Sensor nodes measure an unknown field of interest. We focus on the problem of jointly optimizing the energy allocation of the energy beacon to different sensors and the data transmission powers of the sensors in order to minimize the field reconstruction error at the sink. In contrast to the line of work that focuses on remote estimation problems under total power constraints  or under wireless power transmission with linear EH models \cite{maiShinIshibashi_2017}, we investigate this problem under non-linear EH models. 



We consider the above resource allocation  problem under two different frameworks:  i) an optimization approach where the energy beacon knows the form of the utility function (i.e. average field reconstruction error), channel state information and the EH characteristics of the devices; hence can directly design  resource allocation strategies using an optimization problem and ii) a reinforcement learning (RL) approach where the energy beacon decides on its strategies adaptively based on the battery level information of the nodes and feedback on the utility function. Recently, deep reinforcement learning techniques have shown state-of-the-art performance in continuous control tasks \cite{duan2016benchmarking} and machine learning in wireless networking applications has been recently investigated \cite{7792374}. Our results illustrate that although optimization and  RL approaches have access to different types of knowledge on the system parameters, they are able to obtain the same error levels in the sensing problem considered here. 


{\emph{Notation:}} We denote a column vector by $\mathbf{a}=  [a_1; \ldots;  a_n] \in \mathbb{C}^{n \times 1}$ where semi-colon $;$  is used to separate the rows.  
%
%
The complex conjugate transpose of a matrix $A$  is denoted by $A^\herm$. 



\section{System Model and Problem Statement}

{\underline{Sensing and Signal Model:}}
There are $n_s$ sensors in the system. At time slot $t$, sensor $i$ obtains  $M$ realizations of the random variable ${x_{t}^i}$ and sends it to the sink using a noisy communication channel. The  aim of the sensing system at time slot $t$ is to estimate the $M$ realizations of the  unknown complex proper zero-mean spatially correlated signal ${\bf{x_t}}$ defined as  $ {\bf{x_t}}=[x_t^1;\ldots, x_t^i;\ldots; x_t^{\ns}] \in  \mathbb{C}^{\ns \times 1}$, with  $K_{\mathbf x_t}=\E[\bf{x_t}  \bf{x_t}^\herm ]$, $P_{x_t} \triangleq \tr[K_{\bf{x_t}}] < \infty$.  
The reduced eigenvalue decomposition of   $K_{\bf{\mathbf x_t}}$ is denoted by $K_{\bf{\mathbf x_t}} =U_t \Lambda_{\mathbf x_t}  U_t^\dagger$  where  $\Lambda_{\mathbf x_t} \in  \mathbb{R}^{s \times s}$ is the diagonal matrix of $s$ non-zero eigenvalues and $U \in \mathbb{C}^{\ns \times s}$ is the matrix of eigenvectors. 
In the sequel, a realization of the random variable $x_t^i$ is denoted with $x_{t,j}^i$ for the sake of clarity whenever needed.

{\underline{ Communications to the Sink:}}
Sensors send their observations to the sink using  a  single cell orthogonal division multiple access (OFDMA) set-up where the spectrum is divided into $\ns$ equal sub-channels  where each sensor is assigned to one sub-channel \cite{huangLarsson_2013}.
%
%
During time slot $t$, $M$  measurements of sensor $i$ is sent to the sink in an uncoded manner as follows
\begin{align}\label{eqn:comm}
 {  y_{t,j}^i }= g_t^i \sqrt{\frac{{p_t^i}}{\sigma_{x_{t}^i}^2}} x_{t,j}^i+ w_{t,j}^i, \quad j=1,\ldots, M
\end{align}
where $\sqrt{p_t^i}\in \mathbb{R}$,  denotes the power amplification factor adopted by  sensor $i$ at time slot $t$, $g_t^i \in \mathbb{R}$ is the effective channel gain,  $y_{t,j}^i \in \mathbb{C}$ denotes the received observation and $w_{t,j}^i \in \mathbb{C}$  denotes the zero-mean proper white channel  noise with variance $\sigma_w^2$. 
The channel gain and channel noise variance is assumed to be constant during transmission at  time slot $t$. 
\iftreport
Hence, we choose ${p_t^i}$ values that do not depend on the realization of the random variable.  
\fi
%

The sink collects the measurements from sensors $i=1,\ldots, \ns$ and makes a linear minimum mean-square error (LMMSE) estimate of the unknown values $\xbf_{t,j} =[x_{t,j}^1;\ldots; x_{t,j}^\ns]$, $\forall j$. The resulting average mean-square error for large $M$ is given by   
$
\frac{1}{M}\sum_{j=1}^M || \xbf_{t,j} - \hat{\xbf}_{t,j}||^2 \to \E[|| \xbf_{t} - \hat{\xbf}_{t}||^2 ]
$
where $\hat{\xbf}_{t,j}$ is the LMMSE estimate of the $\xbf_{t,j}$. 
Hence, the estimation error $\err_t(\pbf_t ) \!=\! \E[|| \xbf_{t} - \hat{\xbf}_{t}||^2 ] $  can be written as 
\begin{align}
 \err_t\left(\pbf_t \right) =& \tr \left[(\Lambda_{\mathbf x_t}^{-1} + \frac{1}{\sigma_w^2} U_t^\herm G_t \Pbf_t U_t)^{-1}\right],
\end{align}
where $\Pbf_t \!=\! \diag(\pbf_t) \in \mathbb{R}^{\ns \times \ns}$, $G_t \!=\! \diag(\mathbf{g}_t) \in \mathbb{R}^{\ns \times \ns} $, $\pbf_t \!=\! [p_t^1; \ldots;  p_t^{\ns}] \in \mathbb{R}^{\ns \times 1}$, $\mathbf{g}_t\!=\![|g_t^1|^2/\sigma_{x_{t}^1}^2; \ldots;  |g_t^\ns|^2/\sigma_{x_{t}^{\ns}}^2] \in \mathbb{R}^{\ns \times 1}$ and it is assumed that channel state information $g_t^i$, $\sigma_w^2$ and $K_{\mathbf x_t}$ are known at the sink. 

\begin{figure}
\begin{center}
\includegraphics[width=0.55 \linewidth]{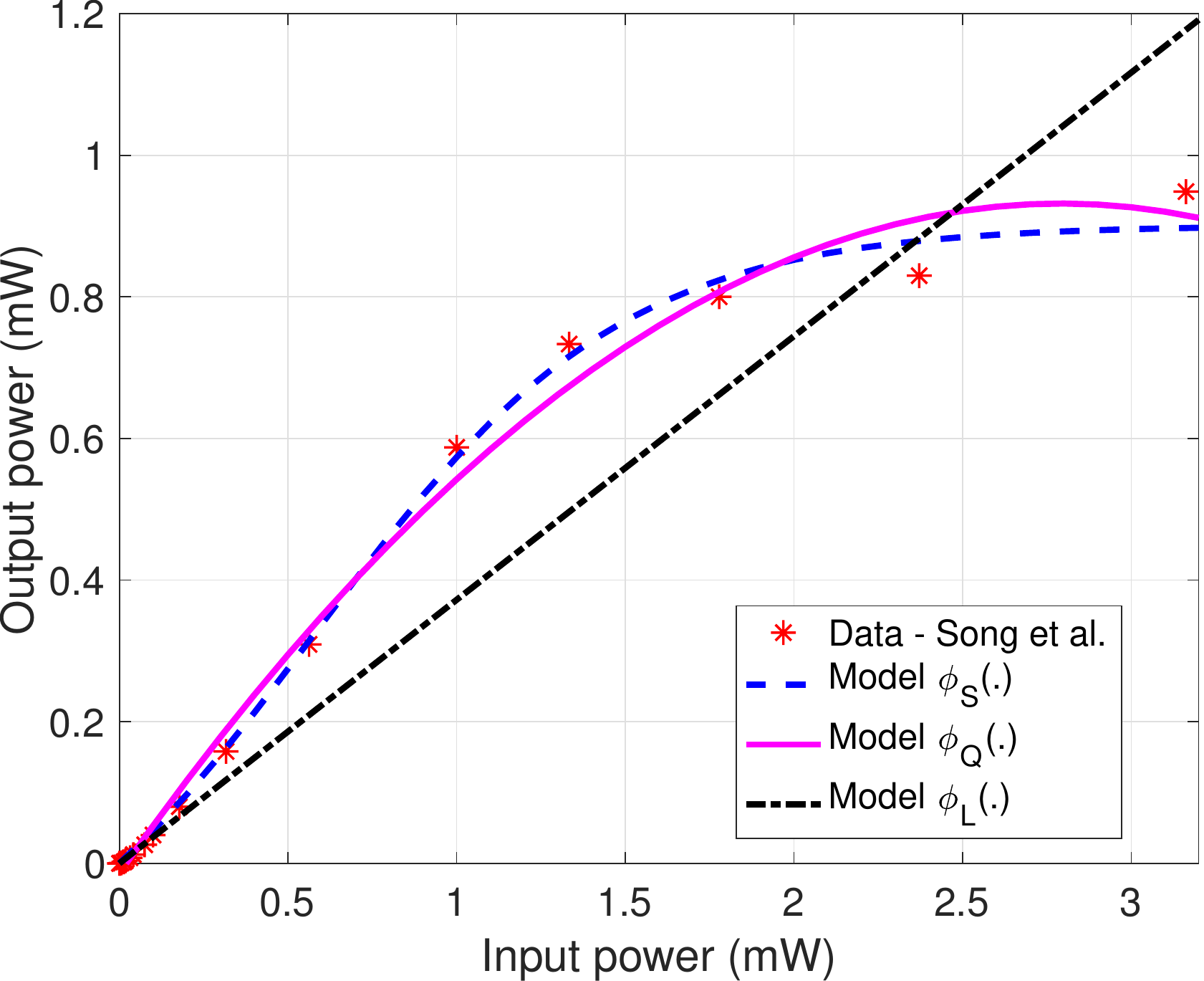}
\end{center}
\caption{Comparison of the measurement data {\protect\cite{SongHuangZhouYuanXuCarter_2015}} and the linear model $\phi_L$, the quadratic model $ \phi_Q$ {\protect\cite{XuOzcelikkaleMcKelveyViberg_2017}},  the logistic function model $\phi_S$ \protect{\cite{BoshkovskaNgZlatanovSchober_2015}}.
} 
\label{fig:nonlineareh}
\kern-0.8em
\end{figure}

{\underline{Wireless Power Transfer:}}
The energy beacon  serves $\ns$ sensors using an orthogonal energy transmission scheme, such as the heterogeneous scenario where devices harvest energy in different frequency bands whereas high EH efficiency in whole spectrum  is challenging to achieve with practical hardware \cite{SongHuangZhouYuanXuCarter_2015}. 
We note that this type of orthogonal energy transmission  formulation also covers energy delivery by time division within time slot $t$ with dedicated sharp energy beams to each sensor \cite{duFischioneXiao_2016}.
The effective channel power gain for power transfer to sensor $i$ during time slot $t$ is denoted by $h_t^i>0$.
The energy beacon allocates an average power  of $q_t^i$ to sensor $i$ at time slot $t$.   
Hence, the  power input to the sensor node $i$ is given by
$
{\bar{q}}_{t}^i = {q_t^i h_t^i}.
$
%
Let the power that can be extracted by the node be denoted by $d_t^i$. The conversion process between ${\bar{q}}_{t}^i$ and $d_t^i$ can be expressed as 
$
   d_t^i=\phi( {\bar{q}}_{t}^i ).
$
where $\phi(.)$ is a possibly non-linear function. 
Hence, the energy harvested by node $i$ during time slot $t$ can be written as 
\begin{align}
E_t^i =\tau_E \phi({\bar{q}}_{t}^i ) =\tau_E \phi(q_t^i h_t^i)
\end{align}
 where $\tau_E$ is the length of energy harvesting time slot.
We consider the following models for $\phi(.)$:
\begin{itemize}
  \item The standard linear model with a constant power conversion efficiency
   \begin{equation}\label{eqn:linear}
   \phi_L( {\bar{q}}_{t}^i ) =\zeta {\bar{q}}_{t}^i,
   \end{equation}
   where $1 \geq \zeta \geq 0$ is the conversion efficiency.  This is the typical model used in the literature  \cite{huangLarsson_2013}.
 \item The quadratic model \cite{XuOzcelikkaleMcKelveyViberg_2017}   
 \begin{align}
 \phi_Q({\bar{q}}_{t}^i )=\alpha_1 {({\bar{q}}_{t}^i)}^2+\alpha_2 {\bar{q}}_{t}^i+\alpha_3,
 \end{align}
 where  $\alpha_1,\alpha_2 ,\alpha_3  \in \mathbb{R}$ are the parameters of the model.
 \item The logistic/sigmoid function model \cite{BoshkovskaNgZlatanovSchober_2015}
 \begin{equation}\label{eqn:logistic1}
\phi_S({\bar{q}}_{t}^i )=\frac{\bar{P}-\beta_3 S}{1-S},
\end{equation}
\begin{equation} \label{eqn:logistic2}
\bar{P}=\frac{\beta_3}{1+\exp(-\beta_1 ({\bar{q}}_{t}^i -\beta_2))},
\end{equation}
where
$
S \triangleq \frac{1}{1+\exp(\beta_1 \beta_2)},
$
 and $\beta_1,\beta_2,\beta_3$ are the  parameters of the model. 
\end{itemize}  
These models are illustrated in Fig.~\ref{fig:nonlineareh}, where the parameters of all the models are found by least-squares curve fitting of the measurement data from the hardware design of \cite{SongHuangZhouYuanXuCarter_2015}.

\underline{Energy Constraints at the Sensors:}
For large $M$, average power consumption  associated with \eqref{eqn:comm} is given by
\iftreport
\begin{align}
 \frac{1}{M} \sum_{j=1}^M  \frac{{p_t^i}}{\sigma_{x_{t}^i}^2} (x_{t,j}^i)^2 \to \frac{{p_t^i}}{\sigma_{x_{t}^i}^2} \sigma_{x_{t}^i}^2 =  p_t^i
\end{align}
\fi
Hence, total energy spent by the sensor at time slot $t$ is given by $J_t= \tau_I p_t^i $, where $\tau_I=M \tau_I^v$ is the duration of the information transmission and $\tau_I^v$ is the average duration of transmission of each sensor value. 
Energy used by the sensor at any time slot could not exceed the available energy. Hence, we have the following energy neutrality conditions
\iftreport
\begin{align}\label{eqn:en}
\sum_{l=1}^t \tau_I  p_l^i \leq \sum_{l=1}^t  \tau_E \phi({\bar{q}}_{t}^i), \quad \quad  t = 1,\ldots,T.
\end{align}
\fi
where the initial energy in the battery is zero and the battery capacity is large enough so that all the energy that is delivered to the device can be stored.

\underline{Problem Statement:}
Our goal is to jointly design the optimal power amplification factors $p_t^i$ at the sensors and energy allocations $q_t^i$ for the energy beacon in order to minimize the mean-square error over the whole time period of $1\leq t \leq T$

\kern-0.1em
\begin{subequations}\label{eqn:opt}
\begin{align}
\label{eqn:objective}
 \,\, \min_{\,\, \substack{\pbf_t, \qbf_t}}  \quad &  \frac{1}{T} \sum_{t=1}^T  \err_t\left(\pbf_t \right)   \\
\label{eqn:cons:eh}
\text{s.t.}  \quad  & \sum_{l=1}^t \tau_I p_l^{i}  \leq \sum_{l=1}^t \tau_E \phi({h}_l^{i} {q}_l^{i}) , \quad   \forall t,  \forall i \\
\label{eqn:totalpow}
& \sum_{i=1}^{\ns} \tau_E {q}_t^{i} \leq P_B, \quad \forall t, \\
 \label{eqn:pos}
 & p_t^i \geq 0,  \quad  q_t^i \geq 0, \quad \quad \forall t,  \forall i
\end{align}
\end{subequations}
where $\qbf_t= [q_t^1; \ldots, ; q_t^\ns] \in \mathbb{R}^{\ns \times 1}$ is the vector of power allocations by the energy beacon at time $t$ and $P_B$ is the power budget of the energy beacon. For notational simplicity, we normalize as $\tau_I=\tau_E=1$ in the rest of the article.

We consider this problem under two different frameworks: In the first approach, we consider this optimization problem, i.e. \eqref{eqn:opt}, directly. Here, the covariance matrix of $\xbf_t$ and all the relevant channel state information (CSI) is assumed to be known. This   off-line optimization set-up serves as a benchmark. 
In the second approach, neither this information nor the form of the objective function is known by the decision maker.
A reinforcement learning approach that uses battery level information at the nodes and feedback on the utility (i.e. distortion) is used to solve this problem. This corresponds to a practical scenario where the sensor nodes and the sink report their battery levels and the utility function to the  decision maker (for instance, at the energy beacon), respectively.
The underlying assumption for the usage of RL approach is that the channels and the statistical properties of the unknown field change in a way that RL agent can learn from the previous experiences and can adaptively form a resource allocation strategy. In Section~\ref{sec:num}, we illustrate this point for the case of periodically changing signal covariance matrix. 

\section{Optimization Approach}
\kern-0.1em
The objective function of \eqref{eqn:opt} is a convex function of $\pbf_t$, since $\tr[X^{-1}]$ is convex over $X \succeq 0$. Whether \eqref{eqn:opt} constitutes a convex optimization problem is determined by \eqref{eqn:cons:eh}. If $\phi({h}_l^{i} {q}_l^{i})$ is a concave function of ${q}_l^{i}$, then the problem becomes convex; since \eqref{eqn:cons:eh} becomes an upper bound on a convex function.  
This is the case for  $\phi_L(.)$ which is linear and hence concave; and for $\phi_Q(.)$ which has $\alpha_1<0$ and hence concave  \cite{XuOzcelikkaleMcKelveyViberg_2017}.
Hence, given a strictly feasible point exists,  the Karush-Kuhn-Tucker (KKT) conditions are necessary and sufficient for optimality. 
In the below, we illustrate the usage of KKT conditions for $\phi_L(.)$ for the special case where the channel gains and the covariance matrix of the field is constant over time but not necessarily over different nodes: 

\begin{prop}\label{prop:optuniform}
Let $|g_t^i|=|g^i|$, $|h_t^i|=h^i$, $K_{\xbf_t} =K_{\xbf} =\diag(\sigma_{x^i}^2)$.  Then, we have the following: 
 i) Let   $\phi=\phi_L$ or $\phi=\phi_Q$  with $\alpha_1 <0$.
 %
%
Optimal $p_t^i$ and $q_t^i$ values do not depend on time.
ii)  Let   $\phi=\phi_L$. Optimal values are given by 
\begin{align}\label{opt:const:lin}
p^i = \left(\sqrt{\frac{1}{\kappa} \frac{h^i\zeta \sigma_{x^i}^2 \sigma_w^2}{ |g^i|^2}}-\frac{\sigma_w^2}{|g^i|^2}\right)^{+}
\end{align}
and $q_i={p^i}/{(h^i\zeta)}$ where  $\kappa > 0$ is a Lagrange multiplier so that  $\sum_i q_i =P_B$ and $c^{+} \triangleq \max(0,c)$. 
\end{prop}
\iftreport
Proof is given in Appendix~\ref{sec:pf:prop:optuniform:report}. 
\fi
Note that,  in the optimal strategy  no energy savings between the time instants occur.

We note that, for $\phi_S(.)$, the problem is in general not convex, and sufficiency of KKT conditions should be further investigated. 
On the other hand, optimal solutions  can be determined using  numerical optimization methods with convergence guarantees for $\phi_L(.)$ and $\phi_Q(.)$  due to convexity \cite{cvx}. 
In Section~\ref{sec:num}, we first solve \eqref{eqn:opt} for $\phi_L(.)$ and $\phi_Q(.)$ using such tools \cite{cvx}. 
We then use the resulting solutions as benchmarks to evaluate the success of the RL approach. Then, we investigate the problem with $\phi_S(.)$ using the RL approach. 

%
\section{Reinforcement Learning Approach}

In this part, we redefine the problem as an RL problem. 
In an RL setting,  the system dynamics is assumed to be Markovian so that the next state of the system depends solely on the current state and the action of the RL agent, i.e. it is independent of the previously visited states. In particular, we assume that both the channel gains and the unknown field are statistically independent random processes over time and the signal covariance matrix changes periodically.
The \textit{agent} aims to maximize a reward signal based on its observations by interacting with the environment without a priori knowledge of transition dynamics of the environment and its rewards. We use the following notation: At step $t$, the \textit{observation} of the agent,  $o_t$, is the limited view of the agent regarding the underlying state of the system. The \textit{action}, $a_t$, is the decision made by the RL agent based on its current observation of the system. The \textit{policy}, $\pi$, guides the decision of the agent by mapping an observation to an action. 
The agent's aim is to reach to an optimum policy which maximizes the sum of discounted rewards over time as given by $G = \sum_t \gamma^t r_t$ where $\gamma$ is defined as the discount factor. Our aim is to minimize distortion, hence we define the reward at time $t$ as the negative of the distortion, i.e. $r_t=-\err_t$, and the RL agent tries to minimize the sum of distortions over multiple time slots. 


We assume that the RL agent can get feedback on the battery levels of the nodes and on the distortion after each step. However, it has no information on the statistics of the field to be estimated; i.e.  $K_{\xbf_t}$.   We consider the observation space of the system as the combination of the energy stored in the batteries of the nodes, $b^i_t$, along with the reward returned in the previous step, $r_{t-1}$. We include the last reward information to capture the current state of the environment. Hence, there are $\ns+1$ observations for the RL agent for an $\ns$-node system. 

RL agent controls both the energy beacon and the sensors. Its actions are 1) energy allocations at the beacon, i.e. $q_t^i$, and 2)  transmission power factors at the sensors, i.e. $p_i^t$. For $q_i^t$, to make sure that the transmitted energy to all nodes equals to the power budget of the energy beacon, $P_B$, as given by (\ref{eqn:totalpow}), we define auxiliary variables $s_t^i$ for each node where the energy transferred to a node is found using an exponential softmax operation 
$
q_t^i = P_B\frac{\exp(s_t^i)}{\sum_j \exp(s_t^j)} 
$
which results in $\sum_i q_t^i=P_B$.

The second action, i.e. $p_t^i$, is limited by the energy stored in the battery of node $i$ at time $t$, $b^i_{t}= \bar{b}^i_{t-1} +\phi({h}_t^{i} {q}_t^{i})$ with $\bar{b}^i_{t-1} \triangleq  \sum_{l=1}^{t-1} \phi({h}_l^{i} {q}_l^{i}) -\sum_{l=1}^{t-1}  p_l^{i}$ as implicitly defined by \eqref{eqn:cons:eh}. The agent only knows $\bar{b}^i_{t-1}$.  We define another auxiliary variable, $0 \leq \rho^i_t \leq 1$, which indicates the ratio of $p_t^i$ to  $b^i_{t}$. Then, the transmission power $p^i_t$ of a node is given by $p^i_t = b^i_{t} \times \rho^i_t$. There are $2n_s$ actions to be determined for an $n_s$-node system.

To represent the policy $\pi$ of the RL agent, we use artificial neural networks due to recent success of  deep neural networks at representing complex policies \cite{duan2016benchmarking}. In our scenario, both state and action spaces are continuous. Hence, RL algorithms for discrete action spaces such as deep Q-learning are not applicable. Naive discretization of the action and state spaces would result in an explosion in the number of states which would make the problem intractable. Hence, here we adopted a policy gradient approach referred as Trust Region Policy algorithm (TRPO) which is suitable for continuous control problems  and has shown state-of-the-art performance in deep RL benchmarks \cite{schulman2015trust,patrick_coady_2018_1183378}. 
\iftreport
Further implementation details are presented in Section~\ref{sec:hyper}.
\fi

\section{Numerical Results}\label{sec:num}


\begin{figure}
\begin{center}
\includegraphics[width=0.7 \linewidth]{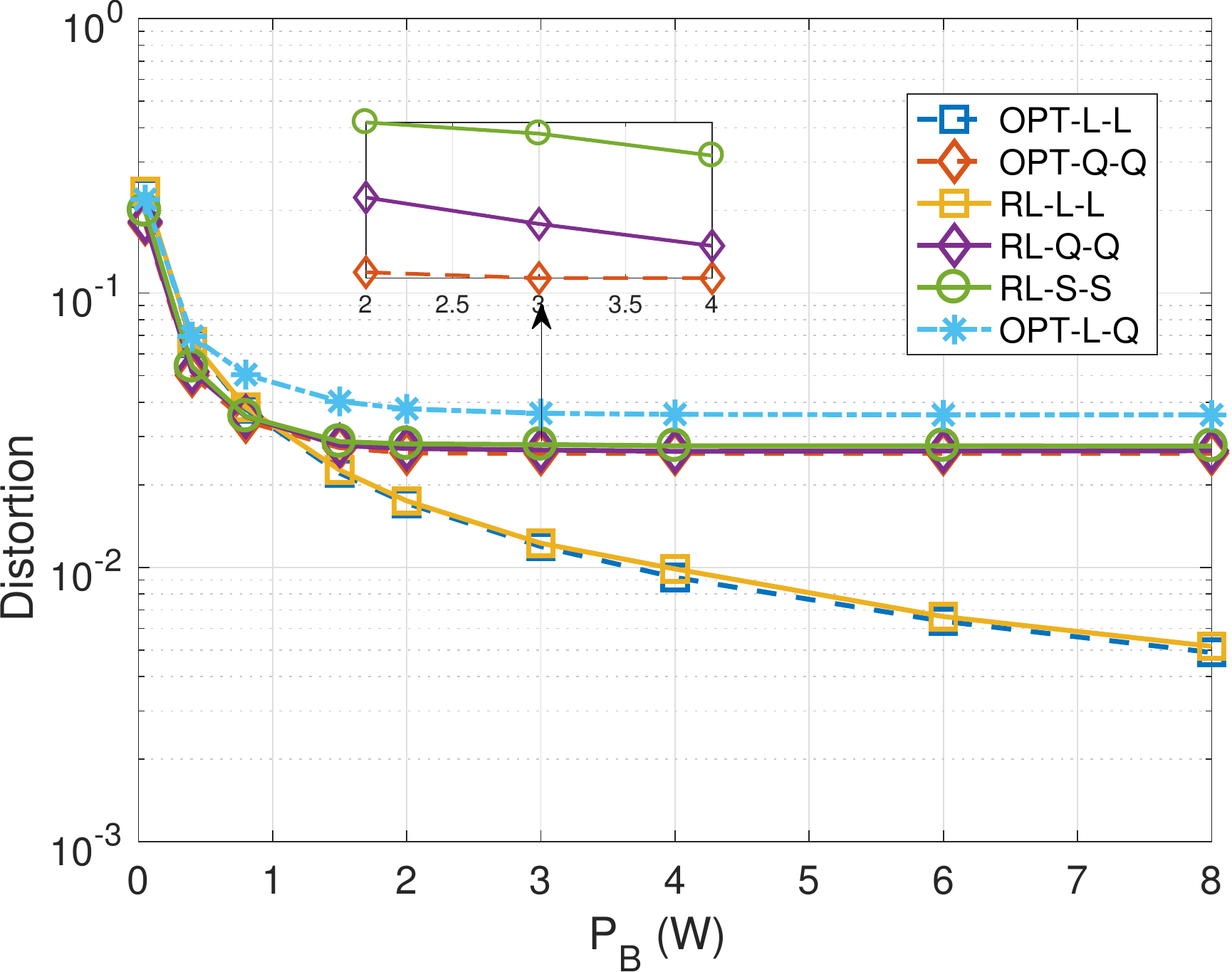}
\end{center}
\caption{Distortion (mean-square error) versus power budget ($P_B$)
}
\label{fig:distortion}
\kern-1.0em
\end{figure}

 \begin{figure}
\begin{center}
\includegraphics[width=0.5 \linewidth]{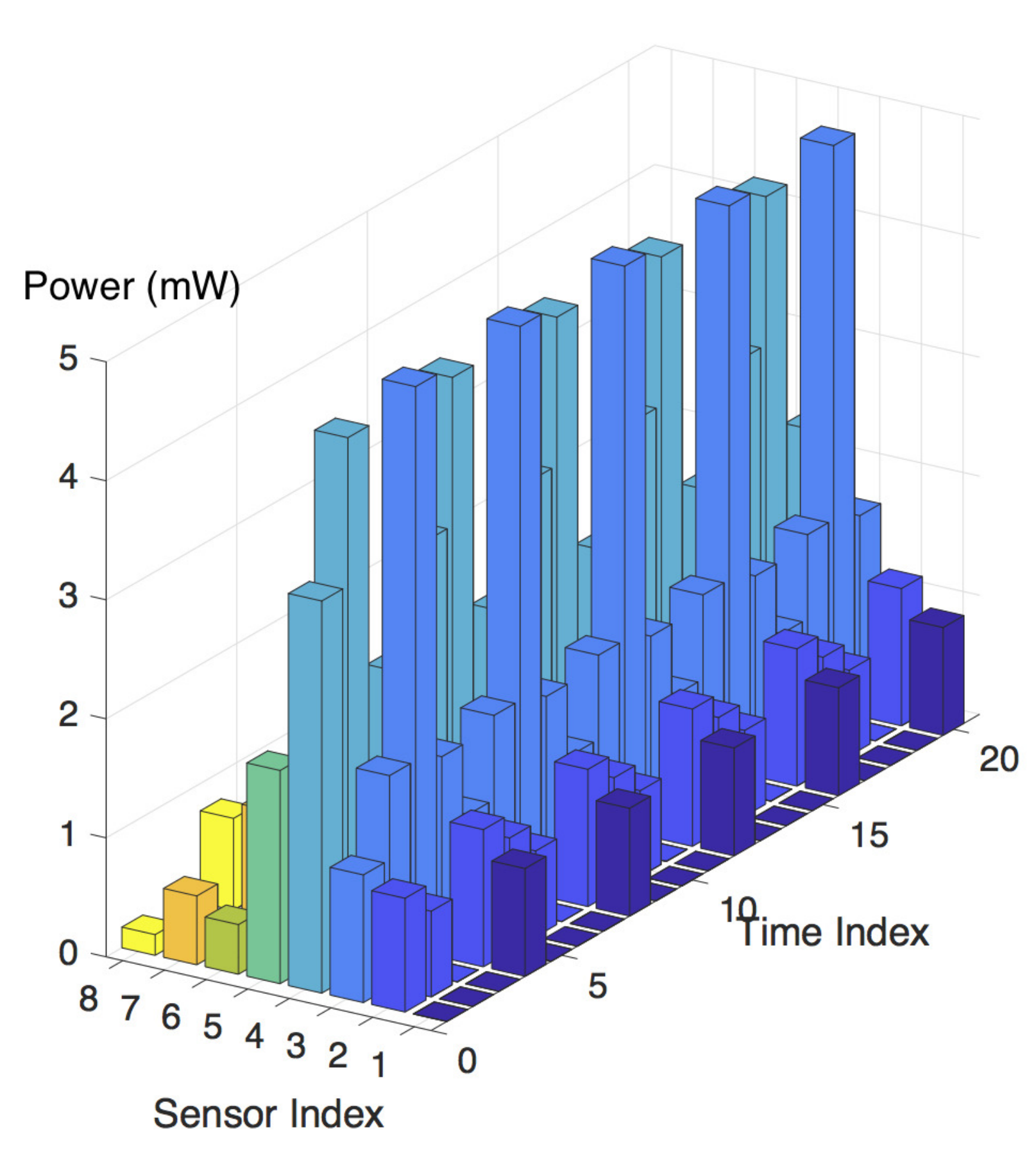}
\end{center}
\kern-1.em
\caption{Information transmission power allocation $p_t^i$, $n_s=8$, $n_t=20$
}
\label{fig:vA}
\kern-1em
\end{figure}


%
Let
$
h_t^i= \frac{A_E A_N}{\lambda^2 {d_i^E}^\gamma_E} |Z_{t,E}^i|^2,
$
where $\lambda$ is the wavelength, $d_i^E$ is the distance between the energy beacon and node $i$,  $A_N$  and $A_E$ are the total apertures of the sensor node and energy beacon  antenna arrays, respectively \cite{huangLarsson_2013}. 
%
The propagation is assumed to be close to line of sight with path loss coefficient $\gamma_E=2$.
Let $f=v_c/\lambda=2.45$GHz, $v_c=3 \times 10^8 m/s$, $A_E=0.2 m^2$, $A_S=0.005 m^2$,  
$Z_{t,E}^i~\sim~\mathcal{CN}(1,0.2)$. 
We have $|g_t^i|^2= \frac{A_I A_N}{\lambda^2 {d_i^I}^\gamma_I} |Z_{t,I}^i|^2$, where $\gamma_I=3$, $A_I =A_E$, $Z_{t,I}^i~\sim~\mathcal{CN}(1,0.2)$ and  $\sigma_w^2 =0.1~\mu$W. Here, $Z_{t,E}^i$ and $Z_{t,I}^i$ are statistically independent of each other and  over $t$ and $i$. 
%
The $d_i^E$ and $d_i^I$ values are set according to the following scenario in 2-D plane: Energy Beacon at  $(\myhyphen 1,0)$, sink at $(4,0)$, node $j$ at $(0,j \myhyphen 4)$,  $j \in \mathbb{Z}$, $j \in [1,8]$ where the unit is meters.  
We assume that  the hardware design of \cite{SongHuangZhouYuanXuCarter_2015} is used for the energy harvesting circuitry. 
%
The second order statistics of $\mathbf x_t$ is periodic in time with the period $\kappa\!=\!4$: $K_{\bf{\mathbf x_t}} \!=\!U_t \Lambda_{\mathbf x_t}  U_t^\dagger$ where   $\Lambda_{\mathbf x_t} =\frac{n_s}{\tr[\Lambda_t]} \Lambda_t$, $\Lambda_t\!=\!\diag(\boldsymbol{\eta}_t)$, $\boldsymbol{\eta}_t=[\eta_{1,t};\dots;\eta_{\ns,t}] \in \mathbb{R}^{n_s \times 1}$ $\eta_{k,t}\!=\!{\nu_t}^k, 0 \leq k \leq n_s-1$, $\nu_t=0.2^{\text{mod}(t,\kappa)}$, where $\text{mod}(t,\kappa)$ denotes modulo operation in base $\kappa$. The unitary matrices $U_t=U_{\text{mod}(t,\kappa)}$ are drawn from the uniform (Haar) unitary matrix distribution.
%
We report the normalized error with $\bar{\err} \in [0,1]$, where $\bar{\err}=\err/P_x$, $\err=\sum_{t=1}^T  \err_t(\pbf_t)$, $P_x=\sum_{t=1}^T \tr[K_{\bf{\mathbf x_t}}]$ and $T=20$.

We label the different scenarios as ``$S\myhyphen AM\myhyphen RM$'' where S $\in \{OPT,RL\}$, AM $\in \{L,Q,S\}$, RM $\in \{L,Q,S\}$ refer to the solution method  (optimization versus reinforcement learning), assumed EH model for optimization and the actual EH model ($\phi_L, \phi_Q,\phi_S$), respectively. For instance, $OPT \myhyphen L \myhyphen Q$ refers to the case where optimization problem in \eqref{eqn:opt} is  solved using the model $\phi_L(.)$ using CVX \cite{cvx} and the performance of the resulting $p_t^i$ and $q_t^i$ values are evaluated based on $\phi_Q(.)$. Hence, this is the scenario where the resource allocation is based on $\phi_L(.)$ whereas the actual EH hardware follows $\phi_Q(.)$. In this scenario, the nodes may not have enough energy to implement $p_t^i$ values found for some time instants due to the erroneous model assumption. For these cases, the node sends with all the energy available. If there is remaining energy, it is used  at $t=T$. It is assumed that energy harvested saturates for input values higher than $2.8$mW for the actual hardware of $\phi_Q(.)$, see Fig.~\ref{fig:nonlineareh}. For RL scenarios, there are no cases with discrepancy between assumed and actual models, since RL makes no assumptions on the EH models and decides on the power allocations based on the feedback on the battery levels and the distortion values.

We first consider the case with $|Z_{t,I}^i|=|Z_{t,E}^i|=1$. The distortion versus power budget $P_B$ curves are presented in Fig~\ref{fig:distortion}. Comparing the RL and OPT curves, we observe that the curves are on top of each other for both $L \myhyphen L$ and $Q \myhyphen Q$ scenarios. This illustrates that RL approach successfully learns how to minimize the distortion even if it does not know the form of this function or the channel gain values. It is also observed there is no significant performance difference between $RL\myhyphen Q\myhyphen Q$ and $RL\myhyphen S\myhyphen S$, which is consistent with the good fit of both models with the measurement data as illustrated in Fig.~\ref{fig:nonlineareh}. Comparing  $OPT \myhyphen L \myhyphen Q$  and $OPT \myhyphen Q \myhyphen Q$, we observe that there is a performance gap due to the wrong assumption on the EH model, which illustrates the need to design resource allocation based on realistic EH models. 



An illustration of the optimal $p_t^i$ values for $OPT\myhyphen L\myhyphen L$ are presented in Fig.~\ref{fig:vA}. The nodes that are closest to the energy beacon and the sink ($j=3,4,5$) are transmitting with the highest power. We observe that nodes save power to be able to transmit with higher power in the subsequent time instants.  We note that the periodic nature of the power allocation scheme over time is consistent with the periodically changing correlation function of the unknown field.

\begin{figure}
\begin{center}
\psfrag{DATADATA1}{ $\phi_L$}
 \psfrag{DATADATA2}{  $\phi_Q$}
 \psfrag{YYY}[bc][bc]{ MSE }
 \psfrag{XXX}[Bc][bc]{  $P_B (mW)$ }
\includegraphics[width=0.5 \linewidth]{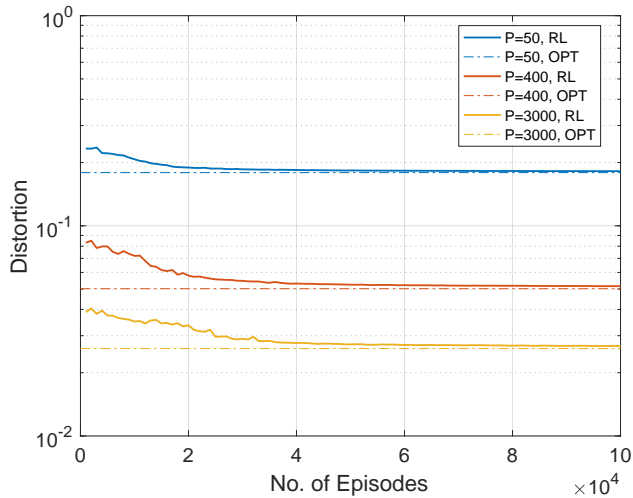}
\end{center}
\caption{Distortion as a function of the number of episodes that the RL algorithm executes. }
\label{fig:RLconvergence}
\kern-1.2em
\end{figure}
Fig. \ref{fig:RLconvergence} shows the convergence behavior of the RL algorithm which also includes the optimum values from the optimization approach as the lower bound. The distortion reduces with more episodes and approaches to the $2$-$3\%$ of the optimum value at the $ \approx 10^5$th episode. Considering that the time horizon is $T=20$ time slots, RL algorithm needs to interact with the system for $\approx 2 \times 10^6$ time slots. 
\iftreport
Further discussions are provided in Section~\ref{sec:disc}. 
\fi

We now discuss the case with $Z_{t,I}^i~\!\sim\!~\mathcal{CN}(1,0.2)$ and $Z_{t,E}^i~\!\sim\!~\mathcal{CN}(1,0.2)$. An average over $100$ channel realizations are reported. We compare the following scenarios: i) Performance of the direct solution of \eqref{eqn:opt} where the CSI for all $t$ are known, ii) Performance of the solution of \eqref{eqn:opt}  for $|Z_{t,I}^i|=|Z_{t,E}^i|=1$ under stochastic channel realizations, iii) Performance of RL which does not know CSI and the form of the  utility function. 
For $P_B = 3W$, we obtain the following normalized distortion values: i) $1.342 \times 10^{-2}$, ii) $3.178 \times 10^{-2}$, iii) $2.89 \times 10^{-2}$
 for  $\phi_L(.)$ and i) $2.73 \times 10^{-2}$, ii) $4.54 \times 10^{-2}$, iii) $4.35  \times 10^{-2}$  for $\phi_Q(.)$. We observe that the distortion values obtained by the RL based approach is reasonably close to these benchmark values. 

\iftreport
\section{Discussions}\label{sec:disc}
We now provide a comparison of the off-line optimization and RL approaches. 
The main advantage of the RL approach against the off-line optimization approach can be considered to be the fact that RL approach does not rely on prior knowledge about the system and learns to optimize performance by interacting with the system.
Applicability of RL approach also does not impose strong necessary conditions on the system set-up other than the Markovian assumption. On the other hand, a standard application of an off-line optimization approach requires full system knowledge, such as CSI information in our set-up. 

Under the off-line optimization approach, for problems that can be formulated as convex optimization problems, well-established numerical methods that guarantee convergence to  an optimal solution can be utilized \cite{boyd}.  On the other hand,  RL approach, in general, does not provide such formal convergence guarantees. Moreover, in practice  RL approach typically requires a  large number of iterations to converge as illustrated in Fig. \ref{fig:RLconvergence}.

Due to this large number of iterations, hence the large number of interactions with the system,   the applicability of RL approach in low-power sensor networks is not straightforward. On the other hand, it should be noted that the off-line optimization approach also requires a significant training overhead since it assumes knowledge of systems parameters, for instance CSI and the covariance matrix of the unknown field in our case. Acquiring this system model information in an accurate manner typically requires multiple interactions with the system. Hence, it is not a priori clear whether the RL approach, which implicitly combines system modeling and optimization, or  the off-line optimization approach, which treats modeling and optimization as separate blocks, is more suitable for sensor network applications. 
We also note that RL approach can be also used in a manner that effectively treats modeling and optimization as separate blocks, i.e. RL agent can interact with a comprehensive system simulation instead of directly interacting with the communication system. Further investigations of these issues are considered as future work. 

\fi

\kern-0.1em
\section{Conclusions}
\kern-0.1em
A comparison of the RL and optimization based approaches for resource allocation in wirelessly powered sensor networks is presented. Practical non-linear EH models are  an important part of the setting.
Our results illustrate that RL based approaches show promising performance with non-linear EH models and partial CSI scenarios. 




\iftreport

\section{Appendix: Proof of Prop.~\protect{\ref{prop:optuniform}}}
\label{sec:pf:prop:optuniform:report}


 i) Since $K_{\xbf} \!=\!\diag(\sigma_{x^i}^2)$,  \eqref{eqn:objective} can be written as $ \sum_{t=1}^T  \sum_{i=1}^\ns \err_i (p_t^i)$, where $ \err_i(p_t^i)= \frac{\sigma_w^2 \sigma_{x^i}^2}{{|g^i|}^2 p_t^i +\sigma_w^2}$.  Suppose that $Q^i \geq 0$ is the total power allocated to node $i$ over the whole time frame, i.e. $\sum_t {q_t^i}\!=\!Q^i$. 
Hence, for node $i$, \eqref{eqn:opt} reduces to: 
\begin{align}\label{eqn:nodei}
\min_{p_t^i} \sum_{t=1}^T \err_i (p_t^i)
\end{align}
such that $\sum_{l=1}^t p_l^i \leq \sum_{l=1}^t \phi(h^i q_l^i)$, $\forall t$ and $\sum_{t=1}^T q_t^i\!=\!Q^i$. We consider the following relaxation of this problem 
\begin{align}\label{eqn:minerrornode}
\min_{p_t^i} \sum_{t=1}^T \err_i (p_t^i)\end{align} such that $\sum_{l=1}^T p_l^i \leq P^i$ where  
\begin{align}\label{eqn:maxpowernode}
P^i= \max_{q_l^i} \sum_{l=1}^T \phi(h^i q_l^i) 
\end{align}
over $\sum_t q_t^i=Q^i$.
(Eqn.~\eqref{eqn:cons:eh} is replaced with one total power constraint and  the right hand side of it  is replaced with its maximum possible value.) 
The objective function of \eqref{eqn:minerrornode} is a Schur-convex function since  $\err_i(p_t^i)$ is convex \cite[Ch.3]{b_marshallOlkin}. Hence, the optimization problem in \eqref{eqn:minerrornode} is a minimization of a Schur-convex function over a total power constraint. Hence, the optimum strategy for \eqref{eqn:minerrornode} is given by uniform $p_i^t$ over $t$  \cite[Ch.3]{b_marshallOlkin}, i.e. $p_i^t=P^i/T$. Similarly, $P^i= T \phi(h^i q_l^i)$ with $q_l^i=Q^i/T$ since \eqref{eqn:maxpowernode} is a maximization of a Schur-concave function over a total power constraint. We now observe that these optimal solutions to \eqref{eqn:minerrornode} and \eqref{eqn:maxpowernode} are also feasible for \eqref{eqn:nodei}. Hence, optimal $p_t^i$ and $q_t^i$ values for \eqref{eqn:nodei} and hence for \eqref{eqn:opt} do not depend on time. 

ii) Using $\phi=\phi_L$ and part (i), i.e. $p_t^i=p^i$ and $p_t^i=q^i$, \eqref{eqn:opt} now can be written as the minimization of $ \sum_{t=1}^T  \sum_{i=1}^\ns \err_i (p^i)=T \sum_{i=1}^\ns \err_i (p^i)$ such that $\sum_{i=1}^\ns q^i = \sum_{i=1}^\ns \frac{p^i}{h^i\zeta} \leq P_B$. The Lagrangian can be written as 

\begin{align}
\mathcal{L}=\sum_{i=1}^\ns \err_i (p^i)+{\kappa} \Omega(p_i)- \mu_i p_i, 
\end{align} 
where ${\kappa} \geq 0$, $\mu_i \geq 0$ are the Lagrange multipliers  and $\Omega(p_i) \triangleq (\sum_{i=1}^\ns \frac{p^i}{h^i\zeta} - P_B)$. 

Solving the KKT conditions, that is differentiating $\mathcal{L}$ with respect to $p_i$ and evaluating it together with ${\kappa}\Omega(p_i)=0$, $\mu_i a_i =0$ and the feasibility conditions,   reveals the solution in \eqref{opt:const:lin}.

\section{Appendix: Hyperparameters of the RL algorithm}\label{sec:hyper}
We approximate the value function with a neural network with three hidden layers having tanh activations. For the 8-node system that we consider, the size of the hidden layers are 90, 21 and 5. The policy is a multivariate Gaussian policy which is also represented with a neural network with three layers having tanh activations. The size of the hidden layers are 90, 127 and 180. We have used Adam optimizer for both networks. We discount future rewards using a  discount factor of $\gamma = 0.995$. 
\fi

\kern1em
 \bibliographystyle{ieeetr}
 \bibliography{JNabrv,WPT} 

\begin{thebibliography}{10}

\bibitem{Krikidis_2014}
I.~Krikidis, S.~Timotheou, S.~Nikolaou, G.~Zheng, D.~W.~K. Ng, and R.~Schober,
  ``Simultaneous wireless information and power transfer in modern
  communication systems,'' {\em IEEE Communications Magazine}, vol.~52,
  pp.~104--110, Nov 2014.

\bibitem{SchoberPoor_2015}
Z.~Ding, C.~Zhong, D.~W.~K. Ng, M.~Peng, H.~A. Suraweera, R.~Schober, and H.~V.
  Poor, ``Application of smart antenna technologies in simultaneous wireless
  information and power transfer,'' {\em IEEE Communications Magazine},
  vol.~53, pp.~86--93, April 2015.

\bibitem{clerckxBayguzinaYatesMitcheson_2015}
B.~Clerckx, E.~Bayguzina, D.~Yates, and P.~D. Mitcheson, ``Waveform
  optimization for wireless power transfer with nonlinear energy harvester
  modeling,'' in {\em 2015 {Int.} {Symp.} on {Wireless} {Comm.} {Sys.}
  ({ISWCS})}, pp.~276--280, 2015.

\bibitem{ClerckxBayguzina_2016}
B.~Clerckx and E.~Bayguzina, ``Waveform design for wireless power transfer,''
  {\em IEEE Trans. on Signal Process.}, vol.~64, pp.~6313--6328, Dec 2016.

\bibitem{huangClerckx_2018}
Y.~Huang and B.~Clerckx, ``Waveform {Design} for {Wireless} {Power} {Transfer}
  {With} {Limited} {Feedback},'' {\em IEEE Trans. on Wireless Comm.}, vol.~17,
  pp.~415--429, Jan. 2018.

\bibitem{Clerckx_2018}
B.~Clerckx, ``Wireless information and power transfer: Nonlinearity, waveform
  design, and rate-energy tradeoff,'' {\em IEEE Trans. on Signal Process.},
  vol.~66, pp.~847--862, Feb 2018.

\bibitem{BoshkovskaNgZlatanovSchober_2015}
E.~Boshkovska, D.~W.~K. Ng, N.~Zlatanov, and R.~Schober, ``Practical non-linear
  energy harvesting model and resource allocation for {SWIPT} systems,'' {\em
  IEEE Comm. Letters}, vol.~19, pp.~2082--2085, Dec 2015.

\bibitem{boshkovska_robust_2017}
E.~Boshkovska, D.~W.~K. Ng, N.~Zlatanov, A.~Koelpin, and R.~Schober, ``Robust
  {Resource} {Allocation} for {MIMO} {Wireless} {Powered} {Communication}
  {Networks} {Based} on a {Non}-{Linear} {EH} {Model},'' {\em IEEE Trans. on
  Commun.}, vol.~65, pp.~1984--1999, May 2017.

\bibitem{XuOzcelikkaleMcKelveyViberg_2017}
X.~Xu, A.~\"{O}z\c{c}elikkale, T.~McKelvey, and M.~Viberg, ``Simultaneous
  information and power transfer under a non-linear {RF} energy harvesting
  model,'' in {\em 2017 IEEE International Conference on Communications
  Workshops (ICC Workshops)}, pp.~179--184, May 2017.

\bibitem{maiShinIshibashi_2017}
V.~V. Mai, W.-Y. Shin, and K.~Ishibashi, ``Wireless {Power} {Transfer} for
  {Distributed} {Estimation} in {Sensor} {Networks},'' {\em IEEE Journal of
  Selected Topics in Signal Processing}, vol.~11, pp.~549--562, Apr. 2017.

\bibitem{duan2016benchmarking}
Y.~Duan, X.~Chen, R.~Houthooft, J.~Schulman, and P.~Abbeel, ``Benchmarking deep
  reinforcement learning for continuous control,'' in {\em International
  Conference on Machine Learning}, pp.~1329--1338, 2016.

\bibitem{7792374}
C.~Jiang, H.~Zhang, Y.~Ren, Z.~Han, K.~C. Chen, and L.~Hanzo, ``Machine
  learning paradigms for next-generation wireless networks,'' {\em IEEE
  Wireless Communications}, vol.~24, pp.~98--105, April 2017.

\bibitem{huangLarsson_2013}
K.~Huang and E.~Larsson, ``Simultaneous information and power transfer for
  broadband wireless systems,'' {\em {IEEE} Trans. Signal Process.},
  pp.~5972--5986, Dec. 2013.

\bibitem{SongHuangZhouYuanXuCarter_2015}
C.~Song, Y.~Huang, J.~Zhou, J.~Zhang, S.~Yuan, and P.~Carter, ``{A
  High-Efficiency Broadband Rectenna for Ambient Wireless Energy Harvesting},''
  {\em {IEEE} Trans. Antennas Propag.}, vol.~63, pp.~3486--3495, Aug 2015.

\bibitem{duFischioneXiao_2016}
R.~Du, C.~Fischione, and M.~Xiao, ``Lifetime maximization for sensor networks
  with wireless energy transfer,'' in {\em Proc. {IEEE} {Inter.} {Conf.} on
  {Communications} ({ICC})}, pp.~1--6, 2016.

\bibitem{cvx}
M.~Grant and S.~Boyd, ``{CVX}: Matlab software for disciplined convex
  programming, version 2.1.'' {http://cvxr.com/cvx}, Mar. 2014.

\bibitem{schulman2015trust}
J.~Schulman, S.~Levine, P.~Abbeel, M.~Jordan, and P.~Moritz, ``Trust region
  policy optimization,'' in {\em International Conference on Machine Learning},
  pp.~1889--1897, 2015.

\bibitem{patrick_coady_2018_1183378}
P.~Coady, ``pat-coady/trpo: First release.,'' Feb. 2018.
\newblock doi: \url{10.5281/zenodo.1183378}.

\bibitem{boyd}
S.~Boyd and L.~Vandenberghe, {\em Convex Optimization}.
\newblock Cambridge University Press, 2004.

\bibitem{b_marshallOlkin}
A.~W. Marshall and I.~Olkin, {\em Inequalities: Theory of Majorization and its
  Applications}.
\newblock Academic Press, 1979.

\end{thebibliography}

\end{document}